\documentclass[superscriptaddress,showpacs,aps,prb,twocolumn]{revtex4}
\usepackage{times}
\usepackage{amsmath}
\usepackage{amssymb}
\usepackage{graphicx}
\usepackage{pst-all}
\begin{document}
 
\title{Identification of the Fractional quantum Hall edge modes by density oscillations} 
\author{Na Jiang}
\author{Zi-Xiang Hu}
\email{zxhu@cqu.edu.cn}
\affiliation{Department of Physics, Chongqing University, Chongqing, 401331, P.R. China}
\date{\today}

\begin{abstract}
The neutral fermionic edge mode is essential to the non-Abelian topological property and its experimental detection in $Z_k$ fractional quantum Hall (FQH) state for $k > 1$.
Usually, the identification of the edge modes in a finite size system is difficult, especially near the region of the edge reconstruction, due to mixing with 
the bosonic edge mode and the bulk states as well.  We study the edge-mode excitations of the Moore-Read (MR) and Read-Rezayi (RR) states by using Jack polynomials
in the truncated subspace. It is found that the electron density, as a detector, has marked different behaviors between the bosonic and 
fermionic edge modes.  As an application, it helps us to identify them near the edge reconstruction and in the RR edge spectrum. 
On the other hand, we systematically study the edge excitations for the RR state, extrapolate the edge velocities and 
their related coherence length and temperature in the interferometer experiments.
\end{abstract}
 \maketitle

\section{Introduction}

Topological states of matter often support gapless edge excitations, which in turn provide a stage to probe its bulk topological properties due to the 
general bulk-edge correspondence.  In the area of the condensed matter physics up to now, there are systems such as integer and fractional quantum Hall
(FQH) liquids, topological insulators and superconductors, et.al.   
The FQH states appear at filling factors with Laughlin sequence have odd denominators in the filling factors due to the Fermi statistics of electrons. 
They are generally addressed by Abelian FQH states since their quasiparticle/quasihole excitations have fractional charge (anyons) and obey the symmetry 
of Abelian braiding group in two dimensions (exchanging two anyons yields an arbitrary phase factor instead of $\pi$ or $2\pi$).  
One exception is the FQH state at $\nu = 5/2$ whose nature are still under debated since its discovery 20 years ago.~\cite{Willett87}
Its most promising candidate ground state wavefunction is a pairing state proposed by Moore and Read~\cite{MooreRead} for a half-filled lowest Landau level
or its particle-hole conjugate.~\cite{FiserPRL07, LevinPRL07, Mtroyer} The MR wavefunction is
\begin{eqnarray}\label{MRWF}
 &&\Psi_{\text{MR}}(z_1, z_2,\cdots, z_N) = \nonumber\\ 
 &&{\rm Pf}\left( \frac{1}{z_i - z_j} \right) \prod_{i<j} (z_i - z_j)^2 \exp\left\{-\sum_i \frac{|z_i|^2}{4} \right\}
\end{eqnarray}
 in which Pfaffian is a $N \times N$ antisymmetric matrix and the quantum Hall analog of a $p+ip$ superconductors in two dimensions.  
 It is the ground state of a three-body $\delta$-function Hamiltonian~\cite{WilczekPRL91} and also a leading candidate 
 ground state for more realistic interactions, such as a Coulomb Hamiltonian with considering the effects of Landau level mixing.~\cite{Mtroyer,Wojs, RezayiPRL11}
 The MR wavefunction is a representative of a universality class which has remarkable properties different from the Laughlin state and their descendents.  
 It supports non-Abelian quasiparticle excitations, with exchanging two 
of them yields a unitary transformation in the ground state degenerate space instead of a pure phase factor as that in the Laughlin state. The ground state degeneracy is 
protected by the topology of its manifold and immune to the environmental perturbation which fuels the interest of the topological quantum computation.~\cite{Kitaev, Freedman, DasSarma}
Including the MR wavefunction, Read and Rezayi~\cite{RR3} constructed a series of non-Abelian quantum Hall states at filling fraction $\nu = k/(k+2)$ which are in the description of 
$SU(2)_k$ Chern-Simons gauge theory.  They are referred to as the RR $Z_k$-parafermion states for reasons of the connection between the wavefunctions and conformal field theory.
The cases of $k=1$ and $k=2$ correspond to the Laughlin and MR states respectively.  And the $k=3$ parafermion FQH state is suspected to be the candidate ground state wavefunction
for the state at filling fraction $\nu=13/5$ and its particle-hole conjugate state at $\nu=12/5$.~\cite{xia04} Recent numerical density matrix renormalization group calculation
shows that it indeed capture the essential characteristics of the ground state of the Coulomb Hamiltonian at these filling factors.~\cite{Wzhu, RMong}
The RR state also supports non-Abelian quasiparticle excitations and it was found to be Fibonacci anyons which support a universal topological quantum computation.

In addition to the non-Abelian quasiparticle excitations in the bulk, the low energy  excitations of the non-Abelian FQH states at edge have a branch fermionic mode (Majorana mode for $k=2$ MR state and 
parafermion mode for $k=3$ RR state) in addition to a charged bosonic mode which only exists in Abelian FQH states.~\cite{WenIJMP,Milovanovic}
The existence of fermionic edge mode makes the low-energy spectrum of the non-Abelian FQHE richer and their experimental consequence more interesting.~\cite{FendleyPRL06}
This is also essential for the non-Abelian statistics and its experimental detection such as in the noise and interference 
experiments.~\cite{chamon97, Fradkin, Sarma, stern06, bonderson06, Rosenow, Bishara, Bonderson, Bonderson08, Willett, Willett13} 
Specifically, the Aharonov-Bohm oscillations in the two point-contact Fabry-P\'{e}rot interferometer strongly depend on the non-Abelian 
quasiparticle parity within the interferometer. In order to observe this phenomena, there is a stringent requirement on the size of the experimental setup which
should not exceed the quasiparticle dephasing length.  The dephasing length of the quasiparticles relies on the velocity difference between the  
bosonic and fermionic edge modes. Our previous numerical results based on exact diagonalization~\cite{Xinprb08,huprb09} for MR state up to 14 electrons
indicate that the stark difference in the edge-mode velocities can lead to ``Bose-Fermi separation'' while quasiparticle propagating along the edge. 
In the exact diagonalization calculation, we need to hybridize some percentage of three-body Hamiltonian in the Coulomb Hamiltonian since the strong mixing between
the edge states and the bulk states. For the calculation of the edge velocities, people just needs the edge spectrum with small momentum up to $\Delta M = 2$ . For small
$\Delta M$, it is not difficult to distinguish the fermionic and bosonic edge states since they have stark difference in the edge velocity, or in the 
dispersion relation near $\delta k=0$. Our numerical
diagonalization results for system up to 14 electrons are $v_c \sim 6.2 \times 10^6$cm/s for bosonic mode and $v_n \sim 0.9 \times 10^5$cm/s for fermonic mode.
However, the identification of the edge modes for the larger momentum sector is not that obvious since these two edge modes are mixed with each other again for
finite size system. Therefore, there is a requirement to search a way to identify the bosonic/fermionic edge state for large momentum sector. It would be useful, as we will discuss 
in this paper, to understand the evolution of the edge mode dispersion as smoothing the confinement, i.e., approaching to the edge reconstruction. 
The numerical diagonalization is limited by the exponentially increasing Hilbert with the system parameters, such as the number of electrons and their occupied
orbitals.

The model wavefunctions of the edge excitations can be constructed by
multiplying the corresponding ground state wavefunction by symmetric polynomials.~\cite{WenIJMP,Milovanovic}  The holomorphic part of the wavefunction
can be calculated analytically from the conformal field theory (CFT) approach by a correlator of bulk CFT primary fields and additional edge fields in the 
general chiral CFT algebra.~\cite{Dubail} Alternatively, the ground state and quasihole state wavefunctions for the $Z_k$ RR
states can be produced recursively by the Jack polynomials with its corresponding root configurations and a negative rational parameter $\alpha$.~\cite{bernevig08,bernevig08a}
Since its computation advantage, the Jack polynomials can achieve much larger system size than that in the exact diagonalization. 
Previous numerical calculations~\cite{XinPRB03,Xinprb08} show that the Hilbert space of the edge excitations is robust even in the presence of a realistic 
long-range interaction, when their excitation energies are comparable to the bulk energies.  Based on this fact, we developed a method to construct the edge excitations 
for FQH states by using Jacks.~\cite{Kihoon}  The system size was extended up to 20 electrons for MR state.
The calculations of the edge spectrum and the edge velocities
were consistent with the previous numerical diagonalization,~\cite{Xinprb08, huprb09} and of course, with more accuracy. 
Especially,  we confirm with greater confidence that the neutral mode velocity 
is an order of magnitude smaller than the charge mode velocity in MR state. 
In this paper, we report a way of identifying the edge states for the non-Abelian FQH state. It is found that the density oscillations of the 
bosonic and fermionic edge modes have a dramatic difference which persists for large momentum space and even after edge reconstruction. We use this criterion
to identify the reconstruction of the fermionic edge mode in a case with pure Coulomb interaction and a background confinement for FQH at $\nu = 5/2$. Also, we verify that this criterion of 
the density oscillations is also applicable for RR edge states. We also extrapolate the edge velocities and their experimental related
coherence length and temperature for MR and RR states after stressing the limitation of the calculation.

The rest of the paper is organized as follows.  In the following section, we introduce the model and the method for studying the edge states in FQH states.
In section III, we discuss the effects of the density oscillations for the bosonic/fermionic edge modes. 
The application of the density criterion in the reconstruction for MR state is discussed in section IV.
In section V, we discuss the edge velocities and related coherence length and temperature in the thermodynamic limit for MR and RR states.
Conclusion and acknowledgements are presented in section VI.

\section{Model and methods}
\label{sec:model}

\begin{figure}
 \includegraphics[width=8cm]{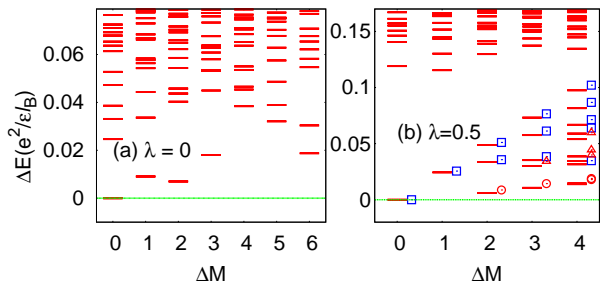}
 \caption{\label{edresult}(Color online) (a) The energy spectrum for $N=12$ electrons with pure Coulomb interaction at 5/2 filling when $d=0.4 l_B$ by exact diagonalization.  
 The global ground state locates in the subspace $\Delta M = M - M_{\text{MR}} = 0$ where $M_{\text{MR}} = N(2N-3)/2$ is the total angular momentum for MR state.  
 (b) The comparision of edge spectrum for $12$ electrons both from ED(plotted with bars) and truncated space (points) in case of $\lambda = 0.5$ and $d= 0.4 l_B$. 
 The fermionic and bosonic edge states are labelled by circle and square points respectively. The triangle points label the mixed states.
 }
\end{figure}

The experimental setup based on semiconductor heterostructure contains the two dimensional electronic gas (2DEG) layer which locates at the interface between GaAs and GaAlAs, and a
uniformly distributed neutralizing background charge attribute to the dopants at a distance $d$.  
Its density is the filling factor $\sigma = \nu$ and the overall charge is the number of the electrons $N$ due to the 
charge neutrality condition. In the case of without considering the Landau level (LL) mixing and the spin degree of freedom, the microscopic Hamiltonian is: 
\begin{equation}
\centering
\label{Hamiltonian}
 H = \frac{1}{2}\sum_{mnl}V_{mn}^l c_{m+l}^+c_n^+c_{n+l}c_m + \sum_m U_m c_m^+ c_m,
\end{equation}
where the $c_m^+$ is the electron creation operator for the lowest LL single electron state with angular momentum $m$ and 
\begin{eqnarray}
\centering
 V_{mn}^l &=& \langle \phi_{m+l}(r_1)\phi_n(r_2)|\frac{e^2}{4\pi\epsilon_0|r_1-r_2|}|\phi_{n+l}(r_2)\phi_m(r_1)\rangle, \nonumber \\
U_m &=& e\sigma \int_{r_2 < R} d^2r_2 \langle \phi_m(r_1)|\frac{1}{\sqrt{d^2 + |\vec{r_1}-\vec{r_2}|^2}}|\phi_m(r_1)\rangle. \nonumber
\end{eqnarray}
Here we work on the disk geometry and $R$ is the radius of the homogeneous background disc, i.e., $R = \sqrt{2N/\nu}$.
The distance $d$ between the 2DEG and the uniform background confinement is our parameter which tunes the relative strength between electron-electron  
repulsion and the attraction from positive background. The interaction matrix element can be routinely calculated in the language of Haldane's pseudopotential and
the background confinement potential can be rewritten in a form of one dimensional integral as follows:
 \begin{eqnarray}\label{um}
U_m = \frac{2Ne^2}{R} \int_0^\infty dk \frac{e^{-dk}}{k} e^{-k^2/2} L_m(\frac{k^2}{2}) J_1(kR),
 \end{eqnarray}
 in which $L_m(x)$ and $J_1(x)$ are the Laguerrel polynomial and Bessel function respectively. However, in the exact diagonalization, we always
fix the number of orbitals, such as 12 electrons in 24 orbitals. The limited $N_{orb}$ attaches another strong edge potential for the edge state in 
subspace with $\Delta M \geq N_{orb} - 2N+2$. In order to eliminate this artificial confinement and mimic a smooth edge circumstance, $N_{orb}$ usually takes a value bigger than $N/\nu$.
Moreover, in the exact diagonalization, we have to deal with a mixed Hamiltonian with a portion of three-body interaction 
 $H_{3B} = - \sum_{i<j<k} S_{ijk} [\triangledown_i^2\triangledown_j^4\delta(\mathbf{r}_i - \mathbf{r}_j)\delta(\mathbf{r}_i -\mathbf{r}_k)]$
 because the strong mixing beween the edge states and the bulk states in the pure Coulomb case, as shown in Fig.~\ref{edresult}(a), obscures the identification of the edge states.
 The MR state in Eq.(\ref{MRWF}) and its edge excitations are exact zero energy
 eigenstates for this model Hamiltonian. Therefore, the edge spectrum of the mixing Hamiltonian 
 $H = \lambda H_{3B} + (1 - \lambda) H_C$ has a well separation from the bulk states~\cite{Xinprb08} as shown in Fig.~\ref{edresult}(b). However, introducing the 
 three-body Hamiltonian makes the Hamiltonian matrix become much more denser which strongly limits and slows down the diagonalization, such as Lanzcos method.  
 Thus later in this section, we will introduce the way of constructing the edge excitations by using the Jack polynomials.

Jack polynomials (Jacks) are homogeneous symmetric polynomials specified by a rational parameter $\alpha$ and a root configuration.  They satisfy a number of differential equations~\cite{Feigin02} and 
exhibit clustering properties.~\cite{bernevig08, Feigin03} For example, Jack is one of the polynomial solutions for Calogero-Sutherland Hamiltonian:
\begin{eqnarray}\label{CSmodel}
 H_{CS}^\alpha=\sum_i (z_i\frac{\partial}{\partial_i})^2 + \frac{1}{\alpha}\sum_{i<j} \frac{z_i + z_j}{z_i - z_j}(z_i\frac{\partial}{\partial_i} - z_j \frac{\partial}{\partial_j}).
\end{eqnarray}
It was found~\cite{bernevig08, bernevig08a} that the FQH wavefunctions for RR $Z_k$-parafermion states can be exactly calculated recursively according to Eq.(\ref{CSmodel}) with a negative parameter $\alpha$ and a root configuration 
(or partition).  The choice of the  root configuration satisfies  $(k, r)$ admissibility which means there can be at most $k$ particles in $r$ consecutive orbitals. The parameter $\alpha$  is $ -(k+1)/(r-1)$ and the corresponding
filling factor is $\nu = \frac{k}{r}$ for bosonic system ($\nu = \frac{k}{k+r}$ for fermionic system, the difference between the  fermionic and bosonic wavefunction is just a Vandermonde determinant). 
For example,  the Jack with $k = 2, r = 2$ ($\alpha = -3$) is the MR wavefunction at $\nu = 1$, which has root ``$20202\cdots$'' in bosonic case and $\nu = 1/2$, root ``$1100110011\cdots$'' in fermionic case.

The edge excitations are  density fluctuation of the incompressible FQH state move along the edge. Their trial wavefunctions can be expressed as a symmetric polynomial multiplied with the ground
state wavefunction Eq.(\ref{MRWF}) as
\begin{eqnarray}
\centering
 \Psi(\{z\}) = S(\{z\}) \Psi_{\text{MR}}(\{z\}),
\end{eqnarray}
where 
\begin{equation}
 S_q(N) = \sum_{i_1<i_2<\ldots<i_q\leq N} z_{i_1}z_{i_2} \ldots z_{i_q}.
\end{equation}
In disk geometry, the total angular momentum for ground state of the fermionic MR state is $M_0 = N(2N - 3)/2$.  Attaching the symmetric polynomials increases the total angular momentum which reflects  the chirality
of the edge mode.  Because of the appearance of the fermionic mode, the number of edge state for MR FQH liquid is $n(\Delta M) = 1,1,3,5,10, 16, 28 \cdots$ for $\Delta M = M_{tot} - M_0 = 0, 1, 2, 3, 4, 5, 6 \cdots$. 
In contrast to  the Laughlin state, the dimension $n(\Delta M)$ of the edge spectrum is generated by 
$\sum_{\Delta M} n(\Delta M)x^{\Delta M} = \prod_{q\geq 1} \frac{1}{1-x^q}$ which are 
$n(\Delta M) = 1,1,2,3,5,7,11 \cdots$ for $\Delta M = M_{tot} - M_0 = 0, 1, 2, 3, 4, 5, 6 \cdots$.
For each angular momentum subspace of a given $\Delta M$, it is easy to label the  basis of the edge spectrum by using  the root of its corresponding Jacks, although they are not orthogonal with each other. 
Such as in the subspace of $\Delta M = 2$ for 8 electrons, there are three Jacks, labelled by 
$|1\rangle = |1100110011001001\rangle$,  $|2\rangle = |110011001100011\rangle$ and $|3\rangle =|110011001010101\rangle$.  After Smidt orthogonalization procedure, the completed basis of $\Delta M = 2$ edge space
has been obtained, namely $|\hat{1}\rangle, |\hat{2}\rangle$, and $|\hat{3}\rangle $. The rest of work is just diagonalizing a $3\times3$ edge Hamiltonian:
\begin{equation}
 H_{\text{edge}}(\Delta M=3) = \left( \begin{array}{ccc}
	\langle \hat{1}|H|\hat{1}\rangle &\langle \hat{1}|H|\hat{2}\rangle & \langle \hat{1}|H|\hat{3}\rangle \\
	\langle \hat{2}|H|\hat{1}\rangle &\langle \hat{2}|H|\hat{2}\rangle & \langle \hat{2}|H|\hat{3}\rangle \\
	\langle \hat{3}|H|\hat{1}\rangle &\langle \hat{3}|H|\hat{2}\rangle & \langle \hat{3}|H|\hat{3}\rangle \\
        \end{array} \right).
\end{equation}
The edge spectrum can be obtained by repeating above procedure in each total angular momentum subspace. To check the completation of the truncated
Hilbert, we span any one of the  edge states $|A\rangle$ from exact diagonalization in the truncated space, labelled by $\{| \hat{k} \rangle \}$
\begin{equation}
 |A\rangle = \sum_k \langle \hat{k}|A\rangle | \hat{k}\rangle.
\end{equation}
Then we define the leakage of the edge space as $L = 1-\sum_k |\langle\hat{k}|A\rangle|^2$ which should be zero if the Hilbert space $\{\hat{k}\}$ is completed.

 For the edge excitation of the MR state, if we assume that each low-energy excitations can be labelled by two sets of occupation numbers $\{ n_b(l_b)\}$ and $\{ n_f(l_f)\}$ for bosonic and fermionic modes,
 and their angular momentum and energies are $l_b$, $l_f$ and $\epsilon_b$, $\epsilon_f$ respectively.  The quantum numbers $l_b$ and $l_f$ are integers and half integers respectively due to the bosonic and 
 fermionic properties.  However, the total fermion occupation number $\sum_l n_f(l_f)$ for each state must be even because each fermionic excitation contains even numbers of Majorana fermion modes due to 
 their pairing nature.  In addition, because of the Pauli's exclusion principle of the fermions, the combinations of one integer angular momentum should be from two different fermionic half integer angular momentums.
 Therefore,  the excitation momentum $\Delta M=2$ can be written as $1/2 + 3/2$ and $\Delta M = 1 = 1/2 + 1/2$ is not allowed. It is easy to figure out that the number of fermionic edge mode are
 $0, 0, 1, 1, 2, 2, 3, \cdots$ for $\Delta M = 0, 1, 2, 3, 4, 5, 6, \cdots$. Including the bosonic mode and assuming they are non-interacting with each other, the angular momentum and energy of the edge states, measured relatively
 to those of the ground state, are~\cite{WenIJMP}
 \begin{eqnarray} \label{lineareq}
  \Delta M &=& \sum_{l_b} n_b(l_b) l_b + \sum_{l_f} n_f(l_f) l_f, \nonumber \\
  \Delta E &=& \sum_{l_b} n_b(l_b) \epsilon_b + \sum_{l_f} n_f(l_f) \epsilon_f.
 \end{eqnarray}
The full edge spectrum, and in addition, the edge velocities can be determined by giving ${\epsilon_b}$ and ${\epsilon_f}$.~\cite{XinPRL02, XinPRB03}
 The results in Fig.\ref{edresult}(b) shows that  the energies of the edge spectrum from the Jacks are consistent with the exact diagonalization with a high accuracy. 
The largest leakage is in the order of $10^{-8}$ for $\Delta M = 4$ and can be furtherly depressed by increasing $\lambda$, or the separation from the bulk states.
The fermionic, bosonic and their mixture are labelled in different point shapes according to 
Eq.(\ref{lineareq}). 
As can be seen, the dispersion of the two edge modes are linear near $\Delta M = 0$ although they have different slopes.
The different dispersion between the bosonic
and fermionic edge states leads to different velocities while quasiparticles propagating along the edge. It induces the ``Bose-Fermi separation'' which is essential to
determine the coherence length and temperature of the quasiparticles as will be discussed in section V.
\begin{figure} 
 \includegraphics[width=8cm]{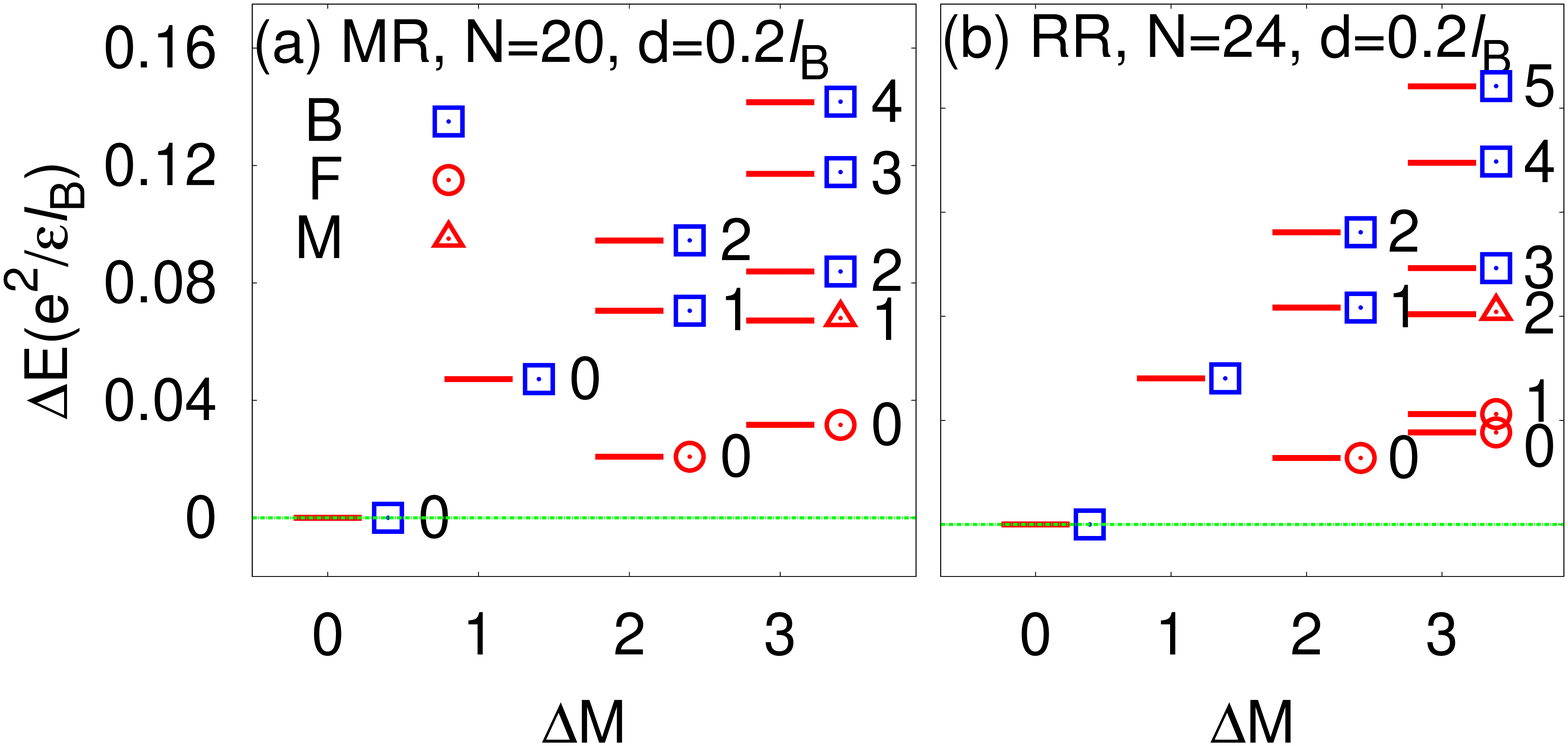}
 \caption{\label{Edgespectrum}(Color online) The edge spectrum(solid bars) for MR state with 20 electrons (a) and RR state with 24 electrons (b) while with background
 potential $d = 0.2 l_{B}$ are from diagonalizing the effective edge Hamiltonian in subspace up to $\Delta M = 3$ . For large system size, as we can see, the fermonic 
 edge states are gapped from the bosonic ones for small $\Delta  M$ which indicates a ``Bose-Fermi'' separation while quasiparticle propagating along the edge.  
 }
\end{figure}

\section{Density differentiation for edge states}
One may say that the energy spectrum fitting from Eq.(\ref{lineareq}) is tricky for large $\Delta M$ since there are many edge states and their energy spectrum is quasi-continuum.
In another aspect, there is a strong mixing between bosonic and fermionic edge mode for a finite size system.
The assumption made in the Eq.(\ref{lineareq}) that the two edge modes are not interacting with each other may not correctly describe the edge spectrum in this case.
On the other hand, determination of the full edge spectrum in $\Delta M$ subspace 
relies on part of the spectrum in $\Delta M + 1$ subspace according to Eq.(\ref{lineareq}).
Therefore, there are arbitrarinesses in identifying the edge modes for large momentum subspace, such as near the region of the edge reconstruction. 

Here an alternative way to distinguish the edge states is contrasting the electron density profiles to that of the ground state.
Intuitively, from the root configuration of the Jacks, since generating the bosonic edge state is generally inserting zeros/fluxes at the outmost edges. 
The fermionic edge state generally appears by a set of consecutive ``1010'' in the root configuration.  
As an example, for system with 10 electrons, we list all the roots of the edge Jacks up to  $\Delta M = 3$ in the following:
\begin{center}
\begin{tabular}{lll}
$\Delta M = 0$ & 110011001100110011  &  \\
$\Delta M = 1$ &11001100110011001\underline{0}1 & B  \\
$\Delta M = 2$  &110011001100110\underline{0}011 & B  \\
 &11001100110011001\underline{00}1 & B \\
 &110011001100\underline{\underline{1010101}} & F \\
$\Delta M = 3$ & 1100110011001\underline{0}10011 & B \\
&110011001100110\underline{0}01\underline{0}1  & B\\
&11001100110011001\underline{000}1  & B\\
&110011001100\underline{\underline{101010}}\underline{0}\underline{\underline{1}}  & M\\
&11001100\underline{\underline{10101010101}}  & F\\
\end{tabular}
\end{center}
Here we label the bosonic (B) component with single underlines and the fermionic (F) component with double underlines. 
The pattern of underline parts are different from the ground state which manifests a perturbation on the edge of the FQH droplet. 
Interesting, the orbital numbers for the two of the most compact Jacks, which are pure bosonic and fermionic edge Jacks, are the same. Such as the root ``1100110011001\underline{0}10011'' and 
``11001100\underline{\underline{10101010101}}'' for $\Delta M = 3$ subspace, both of them occupy 19 orbitals.  Thus in real space, the electrons in these states roughly live in the same area.
However, if we compare these two edge Jacks with ground state at $\Delta M=0$, it is clear that the fermonic mode has a larger affect range than 
that of the bosonic mode in the orbital space, or equivalently, in real space. The above analysis is also applicable to the other bosonic edge Jacks as being indicated from the above root configurations.

Although the edge Jacks are not orthogonal to each other and the real edge states are linear combinations of them as analysis in section II.
The above analysis at least gives us a hint that the fermionic and bosonic edge modes have different perturbation range on the density profile which may be as an evidence 
to distinguish them. Also, Fiete \textit{et.al}~\cite{Fiete08, Fiete10} pointed out that the density fluctuation shows different behaviors while a non-Abelian FQH state coupling to
a nearby quantum dot which can be used as a tool to distinguish the non-Abelian FQH state and its particle-hole conjugate. Here we expect to distinguish the fermionic   
edge states in the edge spectrum via the density fluctuations. As a test, we consider the largest system size for 20 electrons MR state and 24 electrons RR state as we can reach.
The edge spectrum for momentum up to $\Delta M = 3$ is depicted in Fig.~\ref{Edgespectrum}. 
Because of small $\Delta M$, we successfully use Eq.(\ref{lineareq}) to fit the MR edge spectrum and 
label the bosonic and fermionic edge states without ambiguity. 
We label the index of the energy levels for each angular momentum subspace thereupon the 0th state in $\Delta M = 2, 3$ should be fermionic edge
state. For MR edge states in Fig.~\ref{Edgespectrum}(a), the states 1, 2 in $\Delta M = 2$ and 2, 3, 4 in $\Delta M = 3$ are the bosonic edge states.
The state 1 in $\Delta M=3$ is the mixed state which has large overlap with ``$\cdots$110011001100\underline{\underline{101010}}\underline{0}\underline{\underline{1}}''. 
It is shown that in the large system, the fermionic edge states are separated from the 
bosonic edge states by a visible gap for small $\Delta M$. 
For the same analysis, we suspect that the states 1 and 2 in $\Delta M = 2$ and the 3, 4 and 5th states in $\Delta M = 3$ 
subspace for RR edge states in Fig.~\ref{Edgespectrum}(b) are bosonic states since they satisfy
the linear relation in Eq.(\ref{lineareq}) without the fermionic components.  The state 2 in $\Delta M = 3$ 
is Bose-Fermi mixed which is linear combinations of the bosonic and fermionic modes. 
Therefore the counting numbers for the fermionic edge states for RR state are $0, 0, 1, 2 \cdots$ for $\Delta M = 0, 1, 2, 3, \cdots$ which
are different from that in the MR edge spectrum.

\begin{figure}
  \includegraphics[width=8cm]{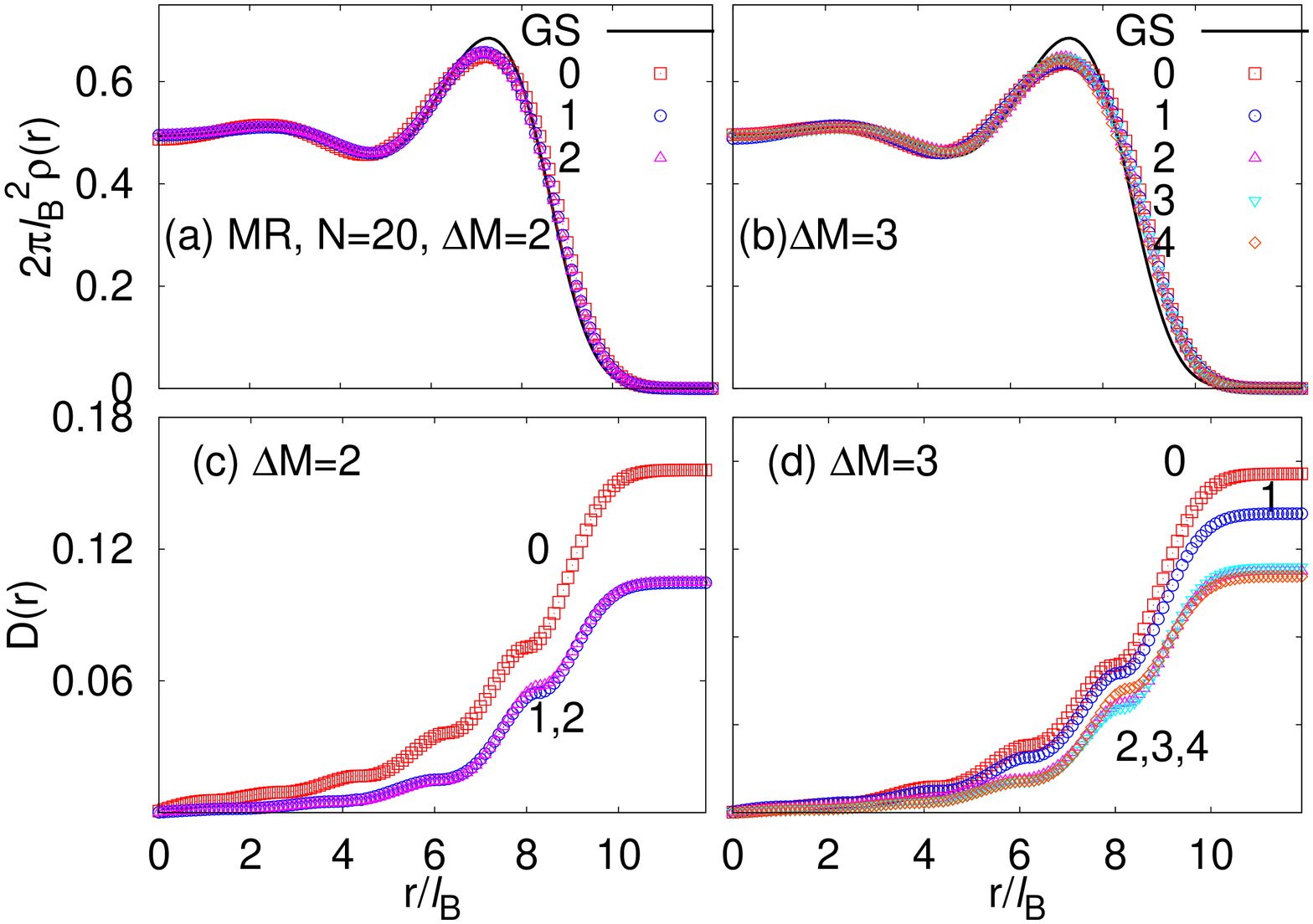}
 \caption{\label{densitydiffMR}(Color online) The radial density profiles for the MR edge states which include the bosonic states, fermionic states, and their mixture 
 in (a) $\Delta M = 2$  and (b) $\Delta M = 3$ subspaces.
 Their corresponding density differentiation $D(r)$
 as a function of $r$ are depicted in (c) and (d). 
 The indexes of the state are one-to-one corresponding to the spectrum in Fig.~\ref{Edgespectrum}(a). 
 }
\end{figure}
 \begin{figure}
\centering
  \includegraphics[width=8cm]{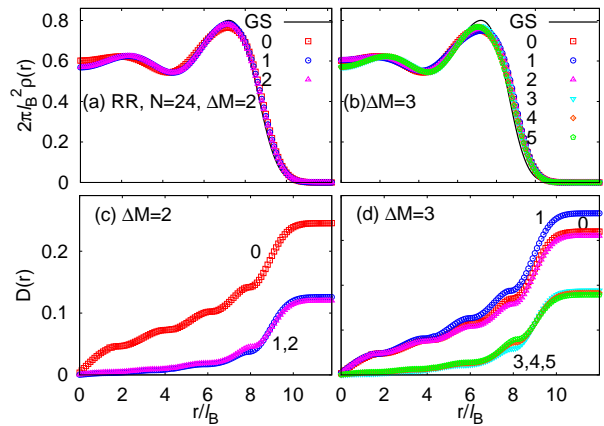}
 \caption{\label{densitydiffRR}(Color online) 
 The same plots as in Fig.~\ref{densitydiffMR} for the RR edge states shown in Fig.~\ref{Edgespectrum}(b).  The $D(r)$ of the fermionic states is still 
 larger than that of the bosonic edge states.
 }
\end{figure}
With these already known edge states, to describe the quantity of the perturbation,
we define the density differentiation between the ground state and the edge state as an integral of their
electron radial density
\begin{equation}
 D(r)=\int_0^r dr'|\rho_{\text{edge}}(r')-{\rho }_{\text{GS}}(r')|.
\end{equation}
Since the edge excitation of the FQH liquid lives on the edge with a finite width in the order of magnetic length $l_B = \sqrt{\hbar c/eB}$, 
it is expected that in the thermodynamic limit, the density differentiation $D(r)$ is zero in the bulk and enhanced as approaching to the edge of the droplet. 
And from above analysis, the final saturated value of the $D(r)$ for fermionic edge state is expected to be greater than that of the bosonic edge state.
The numerical results for MR and RR states are illustrated in Fig.~\ref{densitydiffMR} and Fig.~\ref{densitydiffRR} respectively.
As can be seen from Fig.~\ref{densitydiffMR} (a) and (b), we plot the radial density profiles for all the states of Fig.~\ref{Edgespectrum}(a) in $\Delta M = 2$ and $3$. 
Comparing to the ground state, the density
profiles of the edge states only have observable discrepancy near the edge bump and it is hard to distinguish the fermionic edge states according to this. However, from the 
density differentiation as shown in Fig.~\ref{densitydiffMR} (c) and (d), it is interesting to see that the behaviors of the $D(r)$ for bosonic edge states 
in the same angular momentum subspace have small differences while saturation has been reached. 
Their average saturated value of $D(r)$ is linearly proportional to the angular momentum $\Delta M$. 
As expected, the density differentiation for fermonic edge state is always larger than that of the bosonic edge states which leads to a ``gap'' from the bosonic states while 
saturated.  The saturated value of $D(r)$ for the Bose-Fermi mixed state locates in between that of the pure bosonic and fermionic states which is plausible.
For RR edge states as shown in Fig.~\ref{densitydiffRR}, the results are basically the same except we observing that the $D(r)$ for the fermionic 
edge state has relatively large value while $r \rightarrow 0$ comparing with that in the MR state. It indicates that the parafermionic edge mode
is more extensive than the Majorana fermionic edge mode. 
The nonzero value of $D(r)$ near $r = 0$ for the RR state in our calculation tells us that the center of the system does not
reach a bulk state which should be immune to the edge excitations. Therefore, the numerical calculation for the RR edge state should be more affected by the finite size effect.
This is consistent to the recent iDMRG study the quasiparticle size (about $15l_B$ in diameter which is larger than that of MR state) on cylinder geometry~\cite{RMong}.

\section{Edge modes near reconstruction}
\begin{figure}
\centering
  \includegraphics[width=8cm]{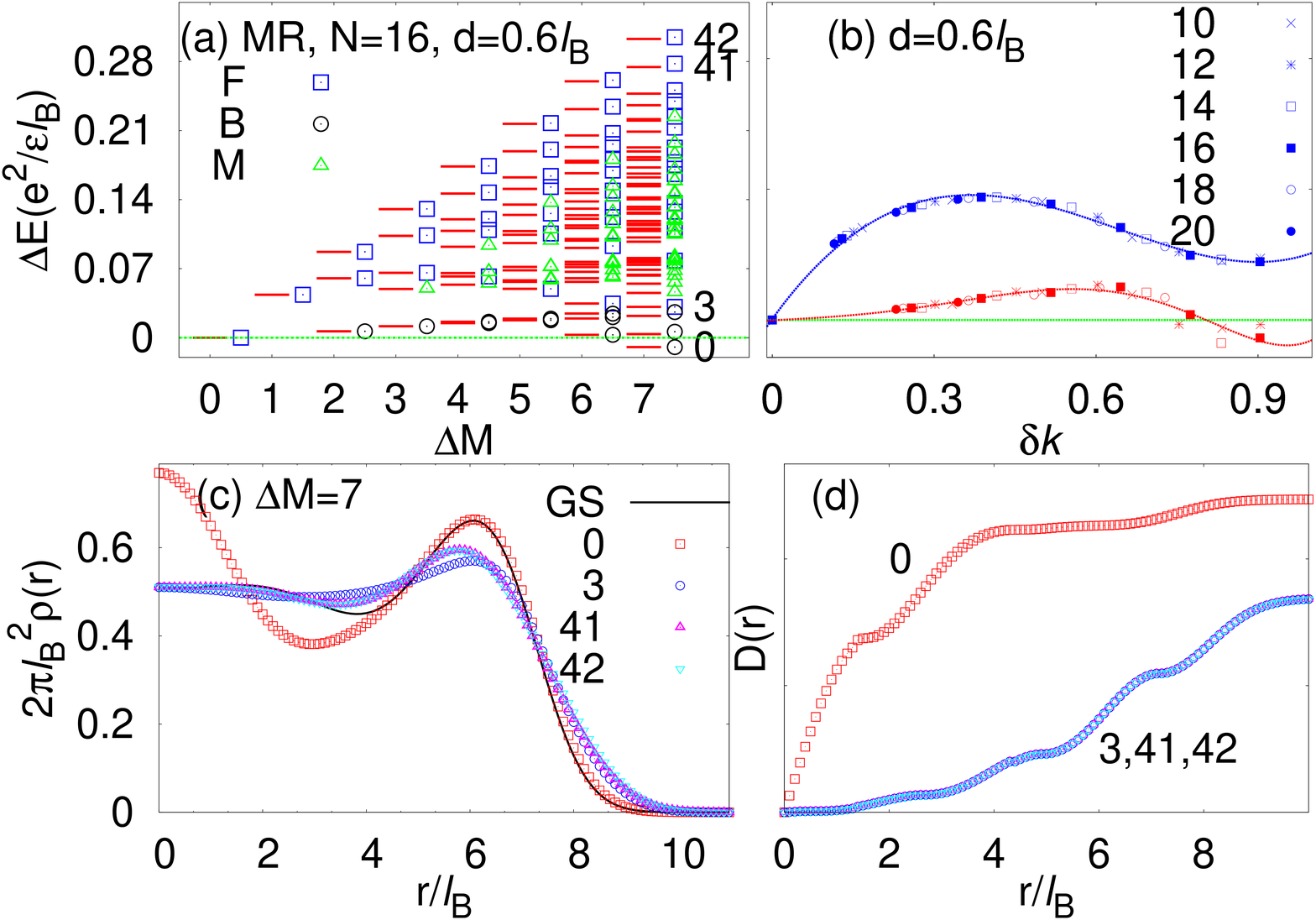}
 \caption{\label{edgereconstruction}(Color online) (a) The edge spectrum for 16 electrons at $\nu = 5/2$ after edge reconstruction.
 The background potential locates at $d = 0.6l_B$ and the first reconstructed state is in $\Delta M =7$.
 (b) The dispersion of the reconstructed edge modes for systems ranges from 10 to 20 electrons. The lower one is fermionic and upper one is bosonic.
 The radial density profiles for the labelled states in (a) and their corresponding density differentiation $D(r)$ as a function of $r$ are shown in (c) and (d) respectively.
 From the density profiles and $D(r)$ in (c) and (d),  we can identify the first reconstructed state is fermionic rather than bosonic.}
\end{figure}

The edge states in FQH liquids are described by chiral Luttinger(CLL) liquid theory by Wen.~\cite{WenIJMP}
It predicts universal properties in FQH droplets, such as the existence of a certain universal electron tunneling exponent for several bulk 
filling factors including the celebrated Laughlin sequence.  For example, CLL theory predicts a power-law current-voltage dependence,
i.e., $I \sim V^\alpha$ in the tunneling between a Fermi liquid metal and a QH edge, where the exponent $\alpha = 3$ for Laughlin state at $\nu=1/3$.
However, such predicted universality has not been observed experimentally in semiconductor based on two dimensional electron gas (2DEG).~\cite{ChangPRL98}  
One possible reason of this discrepancy is electronic density  reconstruction at the edge of FQH liquid.~\cite{Chamon94,XinPRL02}
The edge reconstruction introduces additional non-chiral edge modes that do not correspond to the bulk topology which break down the universality.   
From a simple  electrostatic analysis of the 2DEG in semiconductor heterostructure as did in Ref.~\onlinecite{XinPRB03}, the edge reconstruction can be simply understood as a consequence
of the competition between the positive background charge confinement potential that holds the electrons in the interior of the sample, and the Coulomb repulsion between electrons that
tends to spread out the electron density. Detailed numerical calculations~\cite{XinPRL02,XinPRB03,Jainedge} show that the edge reconstruction occurs quite generally
in the FQH regime, except in some new 2DEG systems, such as graphene.~\cite{HuPRL, EYAndrei} For the FQH state at filling fraction $\nu$ in the thermodynamic limit,  the radius of the droplet is defined as $R_0 = \sqrt{2N_{orb}} = \sqrt{2N/\nu^*}$ where the $\nu^* = \nu - [\nu] = 1/2$ 
is the valence Landau level filling
for $5/2$ state. However, for finite size systems, the number of orbitals $N_{orb}$ has a shift $N_{orb} = 2N-2$ such that $N_{orb} = \phi/\phi_0 = \pi R_0^2/2\pi l_B^2$
and thus $R_0 = \sqrt{4N-4}$. 
To mimic a smooth edge for edge reconstruction, the edge excitations in the angular momentum subspace $ M_{tot} = M_0  + \Delta M$ need at least  $N_{orb} = 2N - 2 + \Delta M$ orbitals, 
or the smallest radius $R = \sqrt{4N-4 + 2\Delta M}$.
 The physical momentum of the edge excitations can be defined as
 \begin{eqnarray}
 \centering
  \delta k \sim  (R - R_0)/l_B^2 \sim \frac{\Delta M}{\sqrt{4N-4}l_B}
 \end{eqnarray}
 in the large $N$ limit.  Therefore, to observe the edge excitation and its reconstruction for large momentum, we need to extend the range of the momentum and also the system size.
 The exact diagonalization can treat system with small size under this demand.~\cite{XinPRL02,XinPRB03} Therefore, we reconsider this problem by the 
 help of the Jacks and the largest system size we have reached is 20 electrons with $\Delta M = 3$ for which the dimension of the Hilbert space is in order of $6\times10^{9}$.
 Based on the data for 10-18 electrons, we find the edge reconstruction occurs at  
 $d_c \sim 0.4l_B$ which is quantitatively agree with the previous study.~\cite{Xinprb08,Jainprb, XinPRL06}
 The lowest metastable state, or the first reconstructed state in the spectrum locates at $\Delta M' = N/2 -1$. 
 Therefore, in order to study the edge reconstruction for $N$-electrons at $\nu = 5/2$ on disk with a smooth edge, 
 the smallest number of orbitals we need is $N_{orb} = 2N-2 + \Delta M' = \frac{5}{2}N -3$.  
 In Fig.~\ref{edgereconstruction}(a), we plot the reconstructed energy spectrum for 16 electrons with $d = 0.6l_B$. The lowest reconstructed state locates in the subspace of $\Delta M = 7$.

With the above criterion of the bosonic/fermionic edge states by the density differentiation, we consider the statistical properties of the lowest reconstructed state
as softening the edge confinement. In Fig.~\ref{edgereconstruction}(c), we plot the density of the 0th state in $\Delta M = 7$. It has a large perturbation comparing with the ground state. This gives us a hint that
the lowest energy state maybe a fermionic state. With this assumption, we find an optimized fitting the spectrum in Fig.~\ref{edgereconstruction}(a) by Eq.(\ref{lineareq}) as displayed
by points. The lowest bosonic state in $\Delta M = 7$ is the 3rd state. We pick out this lowest bosonic edge state and the highest two bosonic (labelled by index 41 and 42) states and 
plot their density and the differentiation comparing with the MR state at $\Delta M = 0$. The results shown in Fig.~\ref{edgereconstruction}(c) and (d) manifest a similar
conclusion as that in previous section.  Therefore, we identify the first reconstructed mode is the fermionic for the $\nu = 5/2$ FQH droplet. With this result, the 
dispersion of the reconstructed edge modes are depicted in Fig.~\ref{edgereconstruction}(b) which shows an anti-crossover behavior or energy level repulsion between the bosonic and 
fermionic modes near the reconstruction momentum. This scenario is consistent to the study of the composite fermion diagonalization~\cite{Jainprb}.

\section{Edge-mode velocities for RR state} 
 Although the existence of the bosonic and fermionic edge modes in the non-Abelian FQH liquid is a univeral property, the edge-mode velocities of them are important nonuniversal
 quantities as they are closely relevant to the coherence length and temperature of the non-Abelian quasiparticle propagating along the edge of the point-contact interferometer.
 Recent numerical DMRG~\cite{Wzhu} and iDMRG~\cite{RMong} calculations show that the bulk properties and their quasiparticle excitations of the FQH states at $\nu = 12/5$ and $\nu = 13/5$
 are captured by the $k=3$ parafermion RR state. To experimentally verify its non-Abelian nature, such as the short noise or point contact experiments,  
 it would be interesting to know more details of the edge physics, such as their propagating velocities along the edge. 

 A non-Abelian quasiparticle carries both a bosonic component and a fermionic component. For the $Z_3$ RR state, the fundamental non-Abelian
 quasiparticle is charged $e/5$ and its operator is $\phi_{qh}^{e/5} = \sigma_1 e^{i\phi_c/\sqrt{15}}$, where $\phi_c$ is the charge bosonic field and $\sigma_1$ is neutral spin field.
 The scaling dimension for $\sigma_{1}$ is $\Delta_{n}=(k-1)/2k(k+2)= 1/15$ and for the vertex operator is
$\Delta_{c}=1/(2k(k+2))=1/30$. Therefore the total scaling dimension is thus $\Delta ^{e/5}=1/10$. 
For completeness, we also consider the related quantities for the Abelian quasiparticle with charge $3e/5$ in RR state.
As known from the CFT, it has scaling dimension $\Delta_{ke/(k+2)}=k/2(k+2) = 3/10$ and no neutral component as that in the non-Abelian case.
For the particle-hole conjugate of the RR state, which may describe the ground state of the FQH at $\nu=12/5$, the scaling dimension 
of the non-Abelian quasihole still contains a charge part $\Delta _{c}^{e/5} = 1/20$
and a neutral fermionic part $\Delta _{n}^{e/5} = 3/20$. As Abelian one, the scaling dimension is $\Delta _{c} = 1/5$.
We also list in Table.~\ref{TableRR} the scaling dimension for the other candidate states of the FQH at filling factor $12/5$ and $13/5$.
The scaling dimensions $\Delta_c$ ($g_c/2$) and $\Delta_n$ ($g_n/2$) play roles in the 
determination of the coherence length $L_{\phi }(T)$ and temperature $T^{*}(L)$ in the double point contact experiment~\cite{Bishara08, Bishara09} as shown in formula from 
the field theory calculation:
\begin{eqnarray} \label{LTformula}
L_{\phi }(T) &=& \frac{1}{2 \pi T}(\frac{g_{c}}{v_{c}}+\frac{g_{n}}{v_{n}})^{-1}, \nonumber\\
T^{*}(L) &=& \frac{1}{2 \pi L}(\frac{g_{c}}{v_{c}}+\frac{g_{n}}{v_{n}})^{-1}.
\end{eqnarray} 
 
 \begin{table}
\[
\begin{array}{|l|c|c|c|c|c|c|}
\hline
\nu = \frac{5}{2}            &   e^{\ast}  &  \text{n-A?}    &   g_c   &   g_n    & L_{\phi}(\mu \text{m})   &  T^{*}(\text{mk})   \\
\hline
\text{MR:}                   &   e/4    &       \text{yes}       &   1/8   &   1/8   &    1.9             &   50.02  \\
                             &   e/2    &    \text{no}        &   1/2   &   0     &     7.17           &  179.13  \\
\hline
\overline{\text{Pf}}\text{:} &   e/4    &   \text{yes}       &   1/8   &   3/8   &    0.70         &  17.49  \\
                             &   e/2    &    \text{no}       &   1/2   &   0     &     7.17           &  179.13\\
\hline
\text{SU$\left(2\right)_2$:} &   e/4    &    \text{yes}          &   1/8   &   3/8   &    0.70         &  17.49 \\
                             &   e/2    &    \text{no}        &   1/2   &   0     &     7.17           &  179.13 \\
\hline
\text{K=8:}                  &   e/4    &    \text{no}           &   1/8   &   0     &    28.66          &  716.53 \\
                             &   e/2    &    \text{no}        &   1/2   &   0     &    7.17           &  179.13  \\
\hline
\text{(3,3,1):}              &   e/4    &    \text{no}       &   1/8   &   1/4   &     1.04          &  25.92 \\
                             &   e/2    &     \text{no}       &   1/2   &   0     &     7.17           &  179.13\\
\hline
\end{array}
\]
\caption{Estimated coherence lengths $L_{\phi}$ at $T=25$mk and coherence temperatures $T^{*}$ for $L=1 \mu$m for the $e/4$ quasiparticles of the candidate $\nu =  5/2$ 
states and the $e/2$ Laughlin-type quasiparticle for all these states. The velocity
estimated in GaAs as $v_{c} \simeq 7.36\times10^6$cm/s and $v_{n} \simeq 5.524\times10^5$cm/s from Ref.\onlinecite{Kihoon}. ``n-A?'' means non-Abelian. $g_c$ and $g_n$
are the charged and neutral scaling exponents respectively. $\overline{\text{Pf}}$ means the particle-hole conjugate of the \text{Pf} state.}\label{TableMR}.
\end{table}

From the above formula, as for the completeness, the edge velocities also play an important role to accurately determine the $L_{\phi }(T)$ and $T^{*}(L)$.
The charge-mode velocity can be defined as $v_{b}=L[E_{0}(\Delta M =1)-E_{0}(\Delta M=0)][e^{2}/\varepsilon \hbar]$
and neutral-mode velocity as $v_{f}=L[E_{0}(\Delta M =2)-E_{0}(\Delta M=0)]/2[e^{2}/\varepsilon \hbar]$. $L$ is the
perimeter of the quantum droplet and $E_{0}(\Delta M)$ is the lowest eigenenergy for the given momentum $\Delta M$.
Based on the results for MR edge excitations up to 20 electrons on disk, we list all the updated coherence length and temperature for all the candidate states
of the FQH at $\nu = 5/2$ in Table.~\ref{TableMR}. 

\begin{figure}
\centering
  \includegraphics[width=7cm]{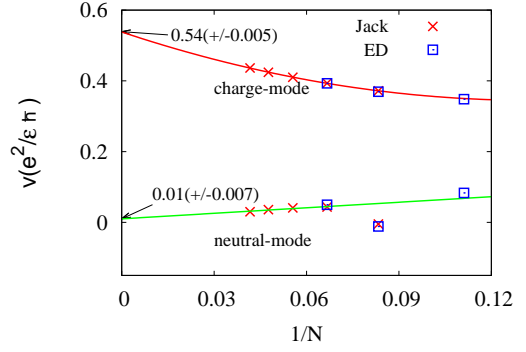}
 \caption{\label{RRscaling}(Color online) Finite-size scaling (9-24 electrons) of the edge-mode velocities for $\nu =
  13/5$ with Coulomb interaction and background confinement $d=0.6 l_{B}$. The charge/neutral mode velocity
  is fitted by a quadratic/linear function of $1/N$ and. In the thermodynamic
  limit, the velocity reads $v_c \simeq 0.54 e^{2}/\epsilon \hbar $ and $v_n \simeq 0.01 e^{2}/\epsilon \hbar$.}
\end{figure}

For the RR state at $\nu = 13/5$, according to the density criterion or the Eq.(\ref{lineareq}) for small momentum, we can also calculate the edge mode velocities
and extrapolate the quantities in the thermodynamic limit. The results are shown in Fig.~\ref{RRscaling}. We use a quadratic and linear functions to fit the data
for bosonic and fermionic mode respectively. The systems which can be
diagonalized in the edge space is extended up to 24 electrons which is much bigger than the 9-15 electrons by exact diagonalization.
The extrapolated values are $v_c \simeq 0.54e^{2}/\varepsilon \hbar$ and $v_n \simeq 0.01e^{2}/\varepsilon \hbar$. 
The charge mode velocity is roughly 6/5 times that of the 5/2 filling, which agrees with the relation $v_c \sim \nu^* e^2/\epsilon\hbar$
from the numerical~\cite{huprb09} and experimental results.~\cite{DTMcClure}
The neutral velocity extrapolated to the thermodynamic limit is very small, in spite of not zero. The velocity of the fermionic mode is roughly 50 times
smaller than that of the bosonic edge mode which indicates a larger ``Bose-Fermi separation'', or smaller coherence length in the interferometer experiments.
These edge velocities correspond to $v_{c} \simeq 9.0\times10^6$cm/s, and $v_{n} \simeq 1.66\times10^5$cm/s in GaAs systems. 
After substituting these velocities into Eq.(\ref{LTformula}) , we estimate the coherence length $L_{\phi }(T)$ 
and temperature $T^{*}(L)$ with the typical experimental temperature $L=25$mk and interference path length $L=1\mu$m.
The results are shown in Table.\ref{TableRR} for all the candidate states for FQH plateau at $\nu = 12/5$ and $\nu = 13/5$.
The coherence length and temperature for the non-Abelian quasiparticles in RR state are roughly $1/3$ of that in the MR state, which means more
stringent requirements are put forward for the experiment for the RR interferometer.

 \begin{table}[t!]
\[
\begin{array}{|l|c|c|c|c|c|c|}
\hline
\nu=\frac{12}{5}(\frac{13}{5})   &   e^{\ast}  &    \text{n-A?}    &    g_c   &  g_n    &     L_{\phi}(\mu \text{m})   &  T^*(\text{mk})  \\\hline         
\text{RR}_{k=3}  &    e/5   &   \text{yes}     &    1/15     &   2/15  &         0.60         &   15.01    \\
          &    3e/5  &   \text{no}      &    3/5      &    0    &         7.30          &  182.54      \\\hline
\overline{\text{RR}}_{k=3} & e/5   &   \text{yes}     &    1/10     &    3/10 &          0.27          &  6.69      \\
         &     2e/5  &   \text{no}      &    2/5      &    0    &         10.95          &  273.81       \\\hline
\text{HH}_{2/5} &     e/5   &   \text{no}      &    1/5      &    2/5  &         0.20         &  5.00       \\
         &     2e/5  &   \text{no}      &    2/5      &    0    &          10.95          &  273.81       \\\hline
\text{BS}_{2/5} &     e/5   &   \text{yes}     &    1/10     &    1/8  &         0.64         &  15.93       \\
         &     e/5   &   \text{no}      &    1/10     &    1/2  &          0.16         &  4.03       \\
         &     2e/5  &   \text{no}      &    2/5      &    0    &         10.95          &  273.81       \\\hline
\overline{\text{BS}}_{3/5}^{\psi}& e/5 &  \text{yes}  &    1/10     &    3/8  &   0.21  &  5.36       \\
         &     e/5   &   \text{no}       &    1/10     &    1/2  &          0.16         &  4.03       \\
         &     2e/5  &   \text{no}       &    2/5      &    0    &          10.95          &  273.81       \\\hline     
\end{array}
\]
\caption{Estimated coherence lengths $L_{\phi}$ at $T=25$mk and coherence temperatures $T^*$ for $L=1 \mu$m for propagating of the Abelian and non-Abelian quasiparticles
along the edge of different candidate ground state for $\nu = 12/5$ and $\nu = 13/5$ FQH states. The velocity estimated in GaAs as $v_{c} \simeq 9.0\times10^6$cm/s and $v_{n} \simeq 1.66\times10^5$cm/s.}
``HH'' and ``BS'' means the Haldane-Halperin~\cite{Haldane83, Halperin84} state and Bonderson-Slingerland~\cite{BS} state respectively.
\label{TableRR}
\end{table}

\section{Summary and acknowledgements}

In summary, we discuss the identification of the bosonic and fermionic edge modes in the non-Abelian MR and RR fractional quantum Hall states. To extend the 
upper bound system size, the truncated edge space spanned by the appropriate Jack polynomials with admissible root configurations is used. As a result, the 20 electrons for MR 
state and 24 electrons for RR state are reached. It is found that the density profile oscillations for the two types of the edge modes have remarkable different influences
on the ground state due to the different width of the edge modes. The fermionic edge mode is usually more expanding than the bosonic one which conduces a large value of the density
differentiation $D(r)$. As an application, this criterion helps us to identify the properties of the lowest reconstructed edge state in the MR edge spectrum as softening
the positive background confinement. The data manifests that the first reconstructed edge mode is fermionic which is consistent to the results by composite fermion diagonalization~\cite{Jainprb,jainbook}
and rencent theoretic discussion of the striped FQH $5/2$ states.~\cite{xin16}

Another application is the RR state for which the structure of the edge spectrum contains parafermionic edge mode. We found the fermionic edge mode in RR state is much more extensive 
than that in the MR state. Thus it suffers more finite size effect. We think this is the reason we did not find a parameter region in disk geometry to stabilize the RR
phase by exact diagonalizing a Coulomb Hamiltonian with confinement potential.~\cite{zhu} We extrapolate the quasiparticle propagating velocities for the bosonic and fermionic edge modes in the thermodynamic limit. 
The related coherence length and temperature in the double point contact interferometer
experiments are listed for all the candidate states for $\nu = 13/5$ and its particle-hole conjugate at $\nu = 12/5$. These physical parameters for MR edge state
are also listed as a supplement for completeness. The coherence length and temperature for the non-Abelian quasiparticles in RR state are roughly $1/3$ of that in the MR state, 
which means more stringent requirements are put forward for the experiment for the RR interferometer.  We hope our results may supply a reference for the future exploring of the identification of the non-Abelian nature of the $\nu=13/5$ 
or $\nu = 12/5$ states, such as the double point interferometer experiments as Willett did in $\nu = 5/2$ state.~\cite{Willett, Willett13}   

 We thank J. K. Jain, X. Wan for useful discussions. This work was supported by NSFC under Project No. 1127403, 11547305 and FRF for the Central Universities No. CQDXWL-2014-Z006.  
 NJ is also supported by CQGSRI under Project No. CYB14033.

\end{document}